\documentclass{elsart}
\usepackage[english]{babel} 
\usepackage{times}
\usepackage[iso]{umlaute} 
\usepackage{graphicx}
\usepackage{cite} 
\begin{document} 
\begin{frontmatter}
\title{Discrete Element Simulation of Transverse Cracking \\During the Pyrolysis of Carbon Fibre Reinforced Plastics to Carbon/Carbon Composites} 

\author[wittel]{Falk K. Wittel\corauthref{cor}},
\corauth[cor]{Corresponding author. Tel.:+49-0711-685-7093 Fax.:+49-0711-685-3706}
 \ead{wittel@isd.uni-stuttgart.de}
\author[2]{Jan Schulte-Fischedick\thanksref{jan}},
\thanks[jan]{Present address:DLR, Institute of Technical Thermodynamics}
\author[3]{Ferenc Kun} 
\author[wittel]{Bernd-H. Kröplin} and
\author[2]{Martin Frieß} 

\address[wittel]{Institute for Statics and Dynamics of Aerospace Structures, University of Stuttgart, Pfaffenwaldring 27, 70569 Stuttgart, Germany}
\address[2]{German Aerospace Center (DLR), Institute of Structures and Design, Pfaffenwaldring 38/40, D-70569 Stuttgart, Germany} 
\address[3]{Department of Theoretical Physics, University of Debrecen, P.O.Box: 5, \\H-4010 Debrecen, Hungary} 

 \begin{abstract}
The fracture behavior of fiber-ceramics like C/C-SiC strongly depends on the initial damage arising during the production process. We study the transverse cracking of the $90^\circ$ ply in $[0/90]_S$ cross-ply laminates due to the thermochemical degradation of the matrix material during the carbonization process by means of a discrete element method. The crack morphology strongly depends on the fiber-matrix interface properties, the transverse ply thickness as well as on the carbonization process itself. To model the $90^\circ$ ply a two-dimensional triangular lattice of springs is constructed where nodes of the lattice represent fibers. Springs with random breaking thresholds model the disordered matrix material and interfaces. The spring-lattice is coupled by interface springs to two rigid bars which capture the two $0^\circ$ plies or adjacent  sublaminates in the model. Molecular dynamics simulation is used to follow the time evolution of the model system. It was found that under gradual heating of the specimen, after some distributed cracking, segmentation cracks occur in the $90^\circ$ ply which then develop into a saturated state where the ply cannot support additional load. The dependence of the micro-structure of damage on the ply thickness and on the disorder in spring properties is also studied. Crack density and porosity of the system are monitored as a function of the temperature and compared to an analytic approach and experiments. 
\end{abstract}
\end{frontmatter}
\newpage

\section{Introduction}

In the recent years extensive investigations were focused on enhancing high-tem-perature performance and reliability for fiber reinforced ceramic matrix composites (CMC), especially their strength, damage tolerance and reliability as structural components. It was found that production process parameters control the formation of the material's micro-structure, and therefore, predetermine the activation of the micro-mechanisms, such as fiber-matrix debonding, fiber bridging and fiber pull-out, leading to damage tolerant behavior \cite{ansorge-95}. Therefore, one goal of ongoing research is to optimize the fiber-matrix interface properties in a way, that the interface is strong enough to avoid excessive fiber pull-out, but weak enough to activate fiber-matrix debonding, in order to avoid a catastrophic failure of the composite.

The present paper focuses on carbon/carbon-silicon carbide (C/C-SiC). This material shows a distinctive crack pattern as a result of the production process, which can be influenced by the laminate layup as well as by the interface properties and the parameters of the production process itself. C/C-SiC is manufactured by the Liquid Silicon Infiltration (LSI) process \cite{krenkel-99,krenkel-luetzenburger-99}, that consists of three major steps. At first a carbon fiber reinforced plastic (CFRP) green compact is produced. This CFRP is then pyrolysed at temperatures ranging from normally 1100 K to 2000 K, converting the thermosetting resin matrix into a glassy carbon one. In this step excessive cracking is observed, driven by the heavy shrinkage of the matrix, which is macroscopically prevented by the carbon fibers \cite{schulte-fischedick-etall-99}. Multiple defects like micro-cracks and micro-delaminations appear. To achieve good mechanical properties in combination with improved abrasive and oxidation resistance of the final CMC, the porous carbon/carbon (C/C) preform is infiltrated with liquid silicon at temperatures higher than 1684 K (melting temperature of silicon). Silicon reacts with both fibers and matrix to silicon carbide resulting in a nearly dense CMC with good overall properties \cite{gern-95,krenkel-fabig-96}. The accessibility of crack systems for liquid silicon during the LSI process predetermines the later constitutive behavior of the C/C-SiC material in a dramatic way. A thermal fiber pretreatment can be employed to influence the interface shear strength and consequently the amount of degraded fibers \cite{krenkel-fabig-96,schulte-fischedick-etall-2002,schulte-fischedick-etall-2001}. In its extreme a siliconization of all fibers leads to a totally brittle behavior of the composite. Therefore, the manipulation of the fiber siliconization is crucial for optimizing C/C-SiC with major influence on all engineering properties as well as on the prediction of the macroscopic failure behavior. Such efforts call for a comprehensive understanding for the development of the crack pattern during the pyrolysis and of the effects of internal structure on the macroscopic response. This paper presents an approach to understand the crack development during the pyrolysis of crossply-CFRP that combines a phenomenological study with a simulation based on a discrete element method. The phenomenological investigations are mainly based on thermo-microscopic observations of the crack development. Additional information was obtained by means of thermo-gravimetry, dilatometry and optical microscopy.
 
Crossply material is built in thickness direction of alternating longitudinal ($0^\circ$) and transverse ($90^\circ$) plies. The mean topic of this study is the simulation of the fracture process during carbonization of crossply material and the characterization of the micro-structure of the occurring damage for different system sizes and disorder. Analytical and numerical predictive methods for the failure of the transverse ply were performed in the past at different length scales ranging from the microscopic scale, modeling single fibers \cite{weihe-etal-94,weihe-kroeplin-93/2} and models for small fiber clusters \cite{asp-berglund-talreja-96,weihe-kroeplin-93,zhu-achenbach-91} up to the meso scale studying homogenized plies \cite{lebon-baxevanakis-etall-98} with a variety of model approaches, but mainly under mechanical loading. Small clusters of fibers with micro-structural disorder, embedded in a matrix material were modeled directly, employing fictitious or discrete crack \cite{weihe-kroeplin-93} models. Unfortunately, these methods are very time consuming and results have limited meaning for the simulation of multiple cracking in thicker transverse plies. In addition, they do not allow predictions about size scaling and the possibility to model the dynamic fracture process itself.

In the presented work, the $90^\circ$ ply is represented by a two-dimensional triangular lattice of springs. The nodes of the lattice model fibers, oriented perpendicular to the plane of the lattice, and springs represent the matrix material in between. The spring-lattice is coupled by interface springs to two rigid bars which capture the two $0^\circ$ plies or adjacent sub-laminates in the model. Disorder is introduced by assigning randomly distributed breaking thresholds to the springs, {\it i.e.} a spring breaks if the load on it exceeds its breaking threshold. A detailed description of the model can be found in \cite{wittel-kun-etal-2001}. To simulate cracking under thermal loading, internal load is imposed on the triangular lattice by varying the initial length of the lattice springs. The time evolution of the system is followed by solving the equation of motion of the nodes (molecular dynamics). With the discrete element method used in this study, realistic system sizes are possible, enabling simulation of the fracture processes with multiple defect interactions of thousands of cracks. Multiple cracking has been studied before, using random spring networks with springs of random distribution of strain to failure \cite{curtin-scher-90,schlangen-garboczi-96,schlangen-garboczi-97}, but rarely in the field of fiber composites research. This approach is particularly suited for studies on dynamic instability in crack propagation, the collective behavior of many interacting cracks and size effects of multiple transverse cracking. Therefore, the discrete element method is practical for studying general features of the statics and dynamics of fracturing, like the crack morphology, global fracture patterns due to the collective behavior of many interacting cracks as well as the dynamic instability in the propagation of cracks. 

Our investigation is focused on the process of damaging, micro-structure of damage, furthermore, on the relative importance of damage mechanisms like segmentation and delamination, driven by the resin thermal degradation during the manufacturing process of the C/C from the CFRP, exposed to a moderate heating environment. Even though our model is capable to predict many aspects of the fracture progress (see \cite{wittel-kun-etal-2001}), we focus on micro-structural properties that can be compared to experimental observations, like the development of segmentation crack density and porosity. Since considerable progress was made in understanding chemical kinetics \cite{trick-saliba-97,trick-saliba-95} and reaction mechanisms \cite{nam-seferis-92/2,wang-96} of the carbonization process, this study is focused on the challenge of combining complex bulk material behavior, composite fabrication and resulting global material behavior. Results from our model can be used for optimization of the production process of all fiber composites that are pyrolysed as well as for the design of functionally graded fiber composites.

It is important to note that the gradual cracking of thin film composites and coatings exhibits certain analogies to the gradual degradation of matrix materials during the production process of composites (\cite{andersons-handge-sokolov-bunde-2000,handge-leterrier-etall-2000,handge-etall-2000,handge-sokolov-blumen-2000}). In these systems a thin film or a so-called coating is attached adhesively to a substrate and the sequential cracking of the overlayer is studied either under mechanical tensile loading or under thermal processing. Especially, the fragmentation of coatings results in segmentation crack scenarios which resemble very much to our system indicating that the underlying microscopic dynamics of cracking is the same in the two cases.

The paper is organized as follows: Section II gives an overview of the carbonization process, describing in-situ observations and measurements. The model construction, tests of the model and the numerical results on cracking obtained by simulations are presented in Section III. An analytic approach worked out in the literature to the thermal transverse cracking problem is briefly summarized in Section IV and confronted with our simulation and experimental results. 

\section{Experimental observations during pyrolysis}
\label{sec:experimentals}

The goal of the phenomenological research is to establish a hypothesis on the crack development during the pyrolysis. This includes three major questions:  \emph{(i)} what types of cracks do occur?  \emph{(ii)} At what stage during pyrolysis do the different types of cracks evolve?  \emph{(iii)} How do the different types of cracks interact? Research was mainly done on laminates of fabric, but for the study of specific questions, experiments were carried out on pure resin, uni-directional laminates and cross-ply laminates. Since the simulation presented here is based on crossply material, this section focuses on this type of reinforcement. 

\subsection{Experimental procedure}\label{sec:experimental-procedure}
Experiments were performed on CFRP specimens made from Tenax HTA fibers and a highly aromatic, nitrogen containing thermosetting resin by resin transfer molding. Uni directional specimens were produced by winding the rovings on a steel core before processing. Crossply material was made by using woven laminates with parts of the weft fibers being removed, reaching a fiber volume fraction of approximately 53\%. All materials were cured at 473 K and post-cured at 510 K before processing them into samples.

Thermo-gravimetric (TGA) and dilatometric analyses were performed to identify the main stages of the pyrolysis as well as to obtain the volumetric changes. The porosity was measured by the water infiltration method \cite{DIN51056} after cooling the specimen from the given temperature. The crack pattern after pyrolysis was observed with an optical microscope with fluorescensic contrast. Note that only cracks contributing to open porosity are highlighted by this procedure. In situ observations were performed under a microscope, equipped with the heating stage (see Ref.\ \cite{schulte-fischedick-etall-99}). The heating rate in all experiments was 10 K/min.

\subsection{Chemistry of the pyrolysis and its influence on mass and volumetric changes}\label{sec:chemistry}

\begin{figure}[htp] \centering{
\includegraphics[scale=0.5]{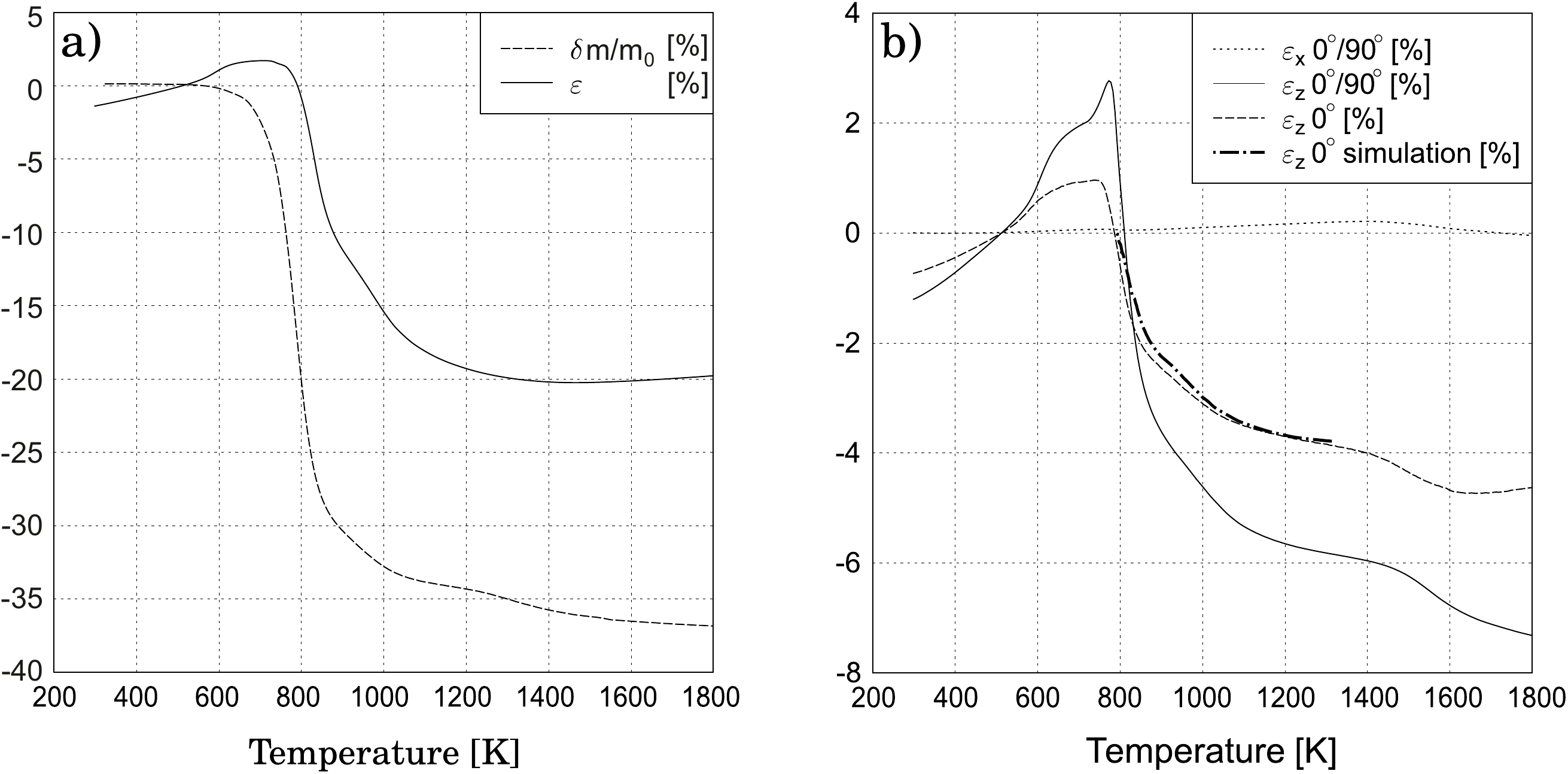}}
 \caption{\label{epsilon_ud}$a)$ Thermo-gravimetric analysis together with dilatometer measurements on resin samples (mean value of nine measurements, three for each specimen orientation). $b)$ Dilatometer measurements on UD and UD cross-ply specimen compared to numerically calculated strain for fiber volume fraction $v_f=0.6$ from section \ref{constcomp}.} 
\end{figure}

Chemical reactions start when the post-curing temperature at 510 K is passed. It can be assumed that the post-curing temperature represents the stress-free state. This is used as a criterion for the starting point of the simulation. Fig.\ \ref{epsilon_ud}$a$ gives an insight in the degradation process by monitoring the mass losses together with the shrinkage of a pure resin sample. In general the pyrolysis can be devided into four stages \cite{jenkins-kawamura-76}: \emph{(i)} During the post-curing stage up to 670 K mainly polyaddition reactions occur with low mass losses and an expansion mainly ruled by the characteristic thermal expansion of the matrix.  \emph{(ii)} Between 670~K and 920~K the main pyrolysis takes place, converting the polymeric matrix into a carbon one with dangling bonds and a decreasing amount of residual hydrogen, resulting in high mass losses, high shrinkages and an extremely high defect density. \emph{(iii)} Up to 1470 K the remaining hydrogen is gradually removed with the observation of only low loss of mass (see Fig.\ \ref{epsilon_ud}$a$). \emph{(iv)} Beyond 1470 K mainly annealing reactions and reactions in the fiber occur \cite{luedenbach-peters-ekenhorst-mueller-98}, leading to further shrinkages in the composite specimens (see Fig.\ \ref{epsilon_ud}$b$). It is important to note that the glassy carbon from the pyrolysis of thermosetting resins is isotropic \cite{jenkins-kawamura-76}. 

%%%%%%%%%%%%%%%%%%%%%%%
%%%%%%%%%%%%%%%%%%%%%%%

Fig.\ \ref{epsilon_ud}$b$ presents the changes in length of unidirectional and crossply samples in the laminate plane ($\varepsilon_x$) and perpendicular to the fiber axis ($\varepsilon_z$). It is important to keep in mind, that all dilatometric data are a superposition of the thermal expansion of the constituents and the thermochemical shrinkages of the matrix (comp. Fig.\ref{epsilon_ud}). It was found that the stiff fibers result in expansion in fiber direction that is one order of magnitude smaller than the out of plane shrinkage. Consequently we neglect the shrinkage in fiber direction during the simulation. Note that the UD specimen can shrink in two dimensions, while the crossply one only has one direction, leading to a higher shrinkage $\varepsilon_z 0^{\circ}/90^{\circ}$. Additionally, a lower fiber volume fraction of the crossply material caused an enlarged shrinkage of $\varepsilon_z$ $0^\circ/90^\circ$ in comparison to the unidirectional $\varepsilon_z$ $0^\circ$. More details can be found in \cite{schulte-fischedick-etall-99}. 

\subsection{Microscopic in situ observations of the pyrolysis}\label{sec:in situ}
\begin{figure}[htb]
\centering{\includegraphics[scale=0.5]{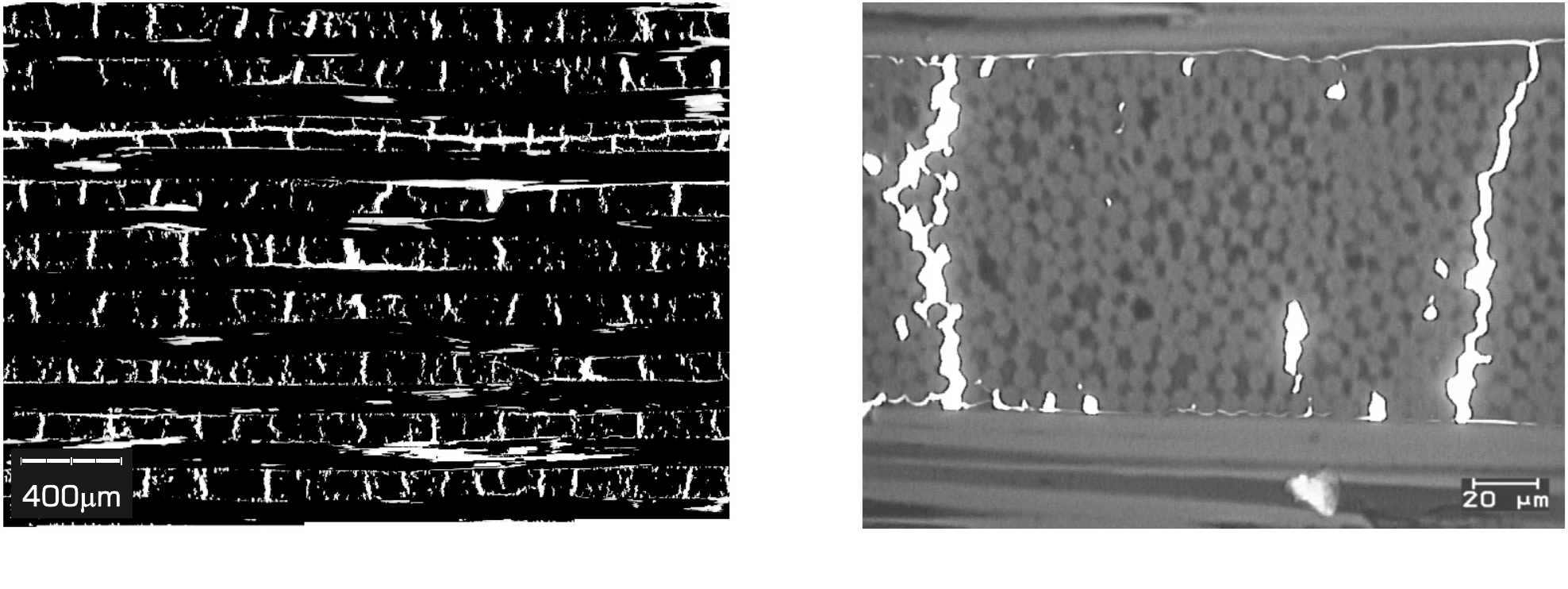}}
\caption{Side view on C/C cross-ply laminate after pyrolysis at 1173 K \label{insitubilder}}
\end{figure}

\begin{figure}[htb]
\centering{\includegraphics[scale=0.5]{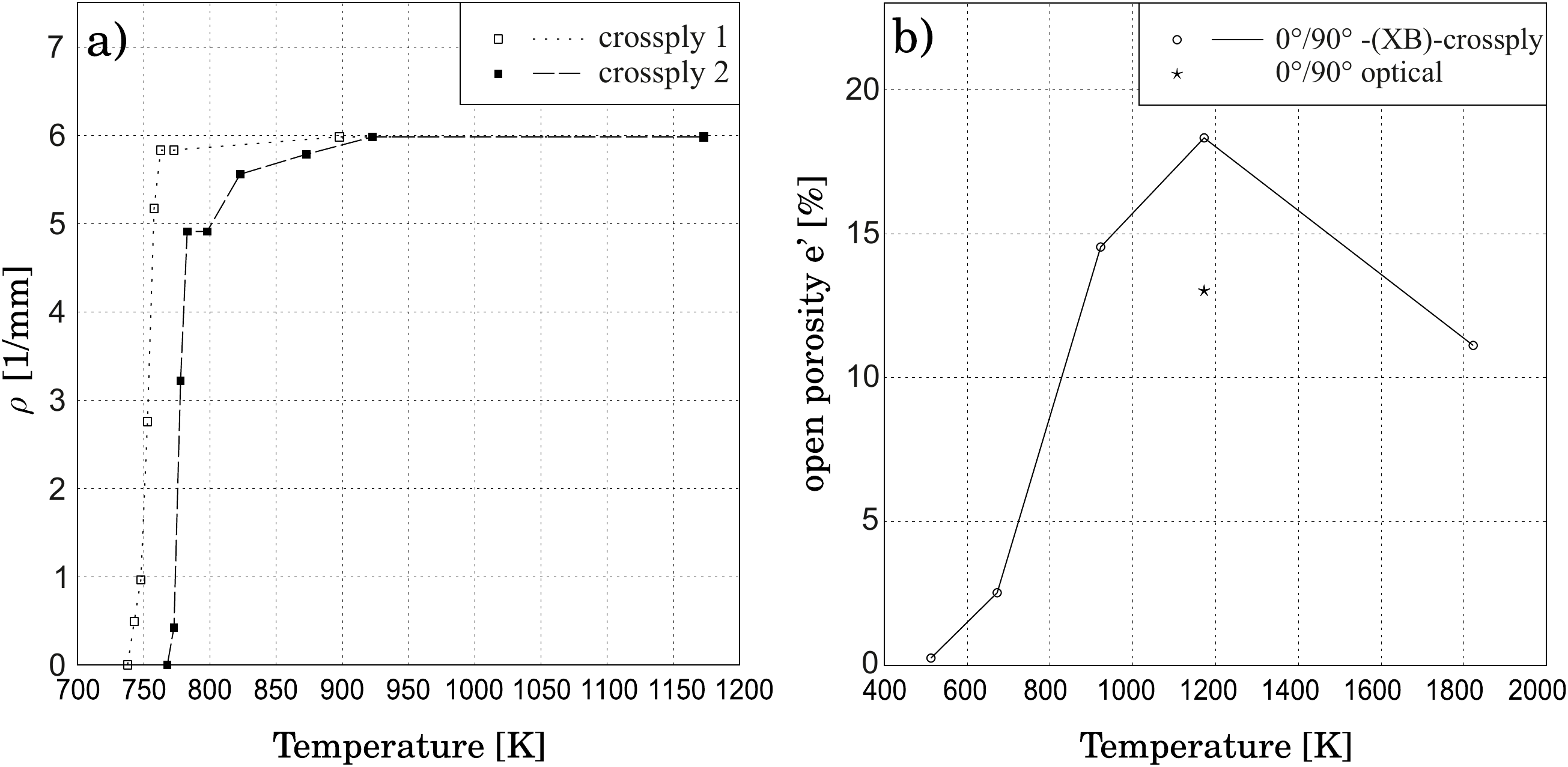}}
 \caption{$a)$ Development of the crack density $rho$ and $b)$ the open porosity $e'$, measured at room temperature. Since the in-situ observation of the development of $\rho$ during the degradation process was limited due to low magnifications of the microscope, we scale these results to the saturation crack density evaluated for the whole specimen after pyrolysis. \label{porosity_crackdensity}}
\end{figure}

The typical micro-structure developing during the pyrolysis is presented in Fig.\ \ref{insitubilder}. Three major crack types, fiber-matrix debonding (fissures without any preferred orientation), transverse cracks and micro-delaminations are observed \cite{gao-patrick-walker-93}.
%%%%%%%%%%%%Insituobservation
%%%%%%%%%%%%%%Harzpyrolyse
The crack development starts with the onset of pyrolysis at approx. 670 K: evolving gases are trapped in the still compact CFRP, so they diffuse into pores leading to a network of fine channels (on the sub-micron scale) through which the gases can escape \cite{nam-seferis-92,wang-96}. The main crack pattern is formed between 750~K and 900~K. First fiber-matrix debondings are observed followed by the coalescence of debondings and pores forming transverse cracks. The development of a crack pattern formed by this crack type is observed within a small temperature range. This can be seen from Fig.\ \ref{porosity_crackdensity}$a$, where the transverse crack density of two thermo-microscopic specimens is shown as a function of the pyrolysis temperature. Micro delaminations evolve from 800~K onward by crack deflection of the segmentation cracks and lead to a saturation of transversal cracking.
%%%%%%%%%%%%%porosity
%%%%%%%%%%%%
As the general rule of cracking during the pyrolysis, a linking mechanism was found: Debondings evolve by linking smaller cracks and pores while transverse cracks and micro delaminations form debonded areas. This mechanism is in good agreement with the results of Sjoergren et al. \cite{sjoegren-berglund-2000} from investigations on mechanically loaded crossply glass fiber reinforced plastics.

The development of the open porosity (see Fig.\ \ref{porosity_crackdensity}$b$) reflects the development of the crack pattern. The main increase takes place during the main pyrolysis in the same temperature range in which the mesoscopic cracks evolve. The maximum of 18\% is reached after a pyrolysis up to 1170 K. Up to now, the mechanisms leading to the decrease of the porosity of 11\% is not understood.

\section{The discrete element model} 
For the study of the formation of defects, a two dimensional triangular spring network model of composites containing rigid cells, is used, that is described in detail in \cite{wittel-kun-etal-2001}. Molecular Dynamics (MD) simulation is used to follow the motion of each cell by solving Newton's equations of motion. In the present study we use a $4^{th}$ order Gear Predictor Corrector scheme \cite{allen-tildesley-87}. A general overview of MD simulations applied to composite materials can be found in \cite{herrmann-roux-90}.
 
This method allows to simulate simultaneously a conglomerate of cracks within rather large lattice sizes with even small cracks being sharply defined. The fundamental advantage of the lattice model used in this investigation is its simplicity, giving direct access and possible physical interpretation to each step of the algorithm. Consequently, one can modify the rules of the model for features of interest, like the characteristic properties of size, strength and of force in a rather straightforward and transparent way.
  
The model is composed in three major steps, namely, {\em (i)} the implementation of the microscopic structure of the solid, {\em (ii)} the determination of the constitutive behavior, and finally {\em (iii)} the breaking of the solid. 

{\em (i) Micro-structure:} 
The model is composed of circular cells of identical radius $r_f$ with their centers located at the nodes of a regular triangular lattice built of sosceles triangles of side length $s$, which is the characteristic lattice spacing. The cells represent the cross sections of fibers which are perpendicular to the model's plane.
%%%%%%%%%%%%%%%%%%%%%%%% 
\begin{figure}[htb] 
 \centering{\includegraphics[scale=0.5]{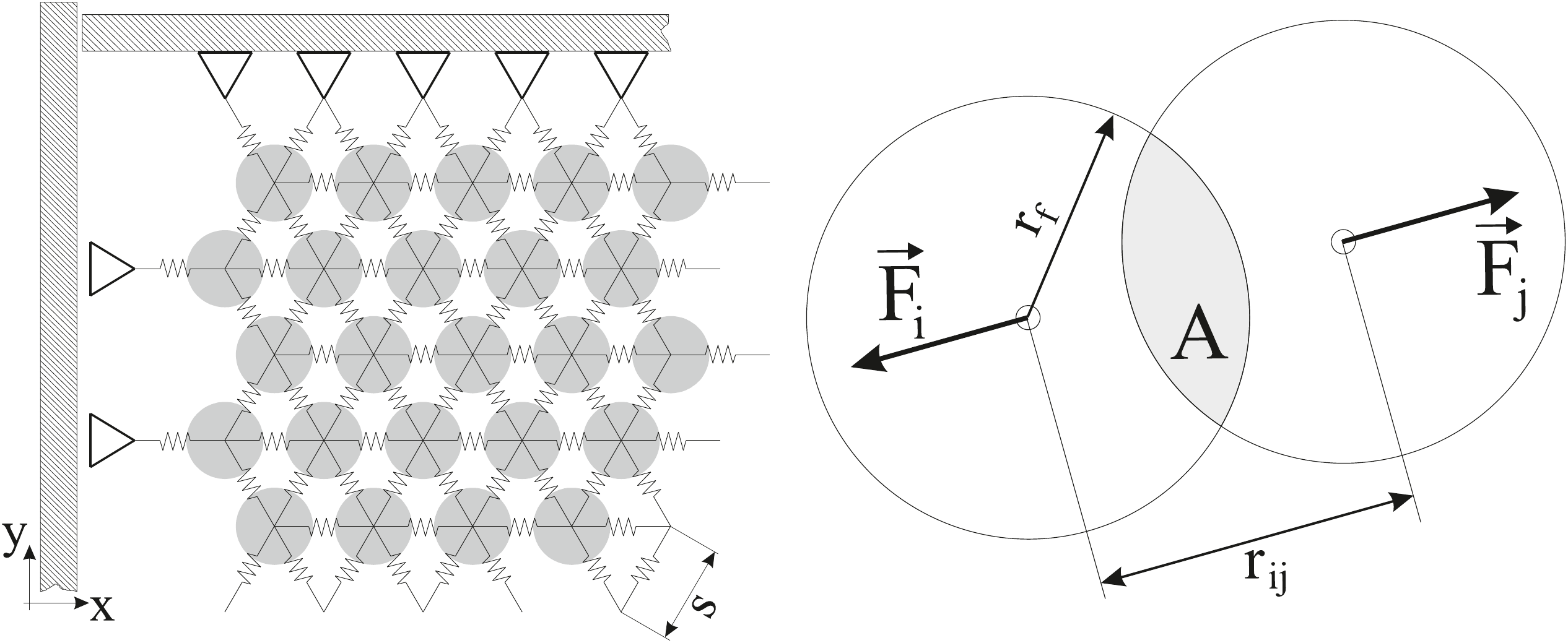}}
 \caption{ \label{microstructure} Micro-structure of the model and cells with contact} 
\end{figure} 
%%%%%%%%%%%%%
The cell centers are connected by linear elastic springs, representing the material region with fiber, fiber matrix interface and embedding matrix material. In our simulation, the cells are the smallest particles interacting elastically with each other. We use a two-dimensional simulation with cells of unit thickness, allowing only motion in the observation plane, with the two degrees of freedom being the two coordinates of the position of the cell center of mass. The utilization of a regular lattice is a clear neglect of the topological disorder, symptomatic for realistic fiber composite systems. 
Since we have no particular basis for a topological disorder, we make a computational simplification by assuming the disorder in spring properties to account for all relevant disorder present in the material. Disorder is given to the model by a Weibull distribution of breaking thresholds $F_d$ at the beginning of the simulation in the form
\begin{equation} \label{weibull}
   P(F_d)=1-\exp\left[-\left(\frac{F_d}{F_0}\right)^m\right] 
\end{equation} 
and is kept constant during the fracture process (quenched disorder). $F_0$ represents the characteristic strength of the material. The Weibull modulus $m$ controls the degree of disorder in the distribution and it is usually chosen in the range $2\leq m \leq 10$, experimentally found to describe a variety of materials. The top and bottom row of cells, as well as the left and right rows are connected to rigid beams with infinite stiffness via interface springs, representing the $0^\circ-90^\circ$ interface of the composite.

{\em  (ii) Constitutive behavior:} 
The fibers are assumed to be rigid, undeformable cylindric bodies with the possibility to overlap when they have contact. This overlap represents to some extent the local deformation of the fibers (soft particle approximation \cite{herrmann-hovi-luding-98}) and leads to a repulsive contact force. The force $\overrightarrow{F}_{i}$ on the cell $i$ is the residual force of all attractive and repulsive forces on the cell calculated from their interaction with neighboring cells. The contact law inside the system is only applied between two circular particles, after the breaking of the connecting spring, thus preventing penetration of broken parts into each other. It is a drawback of this method that the overlap force does not have a Hamiltonian formulation, which implies that no energy can be associated to the deformation modeled by the overlap. Stress is applied to the system by changing the initial length of the springs. The fiber elasticity modulus $E_{ft}$, spring stiffness and breaking threshold is kept constant throughout the simulation.

{\em  (iii) Breaking of the solid:}
Stresses inside the system can be released either by the formation of surfaces (cracks) inside the transverse ply or by fracture of the interface elements on the boundary. In the framework of Discrete Element Methods, the complicated crack-crack interaction is naturally taken into account. If the total force on a spring $F$ exceeds its breaking threshold $F_d$ (damage threshold), its stiffness is abruptly reduced to zero, resulting in a load redistribution to neighboring springs in the next iteration steps. If after some iterations the neighboring springs exceed their threshold value due to the additional load, they fail too, giving rise to crack growth.

The time evolution of the particle assembly is followed by solving the equation of motion of the nodes ({\it i.e.} transverse fibers) 
\begin{eqnarray} 
  \label{eq:eom} 
  m\ddot{\vec{r_i}} = \vec{F_i}, \qquad i=1, \ldots , N 
\end{eqnarray} 
where $N$ denotes the number of fibers of the transverse ply, and $\vec{F}_i$ is the total force acting on node $i$. A $4^{th}$ order Predictor-Corrector algorithm is used in the simulations to solve numerically the second order differential equation system Eq.\ (\ref{eq:eom}). After each integration step the breaking condition is evaluated for all the intact springs. Those springs which fulfill this condition are removed from the further calculations (spring breakage).  

\subsection{Computer simulations} 
Before applying the model to study transverse cracking in fiber composites a variety of simulations have been performed in order to test the behavior of the model and to make parameter identification with respect to experiments. Tests for the Young's modulus, stress distribution between cracks and Poisson's ratio of the model system are described in detail in \cite{wittel-kun-etal-2001}.

\subsubsection{Crack formation under gradual heating} \label{constcomp}
The effect of gradual heating with the resulting shrinkage of the matrix from the degradation process is captured in the model by slowly reducing the initial length of the discrete spring elements, introducing internal stress. The initial length $l^0$ of the spring elements is in the beginning the characteristic lattice spacing $s$ which is determined by the fiber radius $r^f$ and embedding matrix according to Fig.\ \ref{microstructure} as a function of the fiber volume fraction $v_f$
\begin{equation}\label{Kantenlange} 
r^f=\frac{s\sqrt{v_f\sqrt{3}}}{\sqrt{2\pi}}, \qquad  0\leq v_f\leq  v_{f_{max}}=0.906.  
\end{equation} 
The initial length $l^0(T)$ is then calculated with the stress free thermal matrix strain $\varepsilon^m(T)$ 
\begin{equation}\label{Kantenlange2} 
l^0(T)=2r^f+(s-2r^f)(1-\varepsilon^m(T)).
\end{equation}
During the simulation we linearly decrease the strain $\varepsilon^m$. Since the thermal straining of the uni-directional ply in fiber direction is more than one order of magnitude smaller than in transverse direction, we keep the $x$ positions of the interface elements fixed.

The results are presented as a function of the dimensionless strain for the driving matrix shrinkage, giving general validity to the results. Later on we can map the results for comparative reasons to a temperature range of interest, using dilatometer measurements for the pure resin specimen from Fig.\ \ref{epsilon_ud}$a$, starting at approximately 773~K. At this temperature the dilatometric specimens reach the length they had at the tempering temperature. Therefore, it is assumed that the thermal mismatch between longitudinal and transverse ply changes sign. The development of $\varepsilon_z$ from the simulation shows good agreement with measurements in Fig.\ \ref{epsilon_ud}$b$. At 773~K a regular network of pore channels has already developed. As the evolving gases can now escape the material, it is assumed that they do not have any influence on the further crack development and are therefore neglected. TEM investigations on CFRP and C/C-SiC revealed a pore and pore channel diameter of less than 1$\mu$m, which is much less than the characteristic lattice spacing $s$. The effect of the pore channels and small cracks and debondings of this size is therefore not taken into consideration explicitly, but through the disorder of the system expressed by the Weibull-statistics in Eq\ \ref{weibull}.  

\begin{figure}[htb] 
 \centering{\includegraphics[scale=0.6]{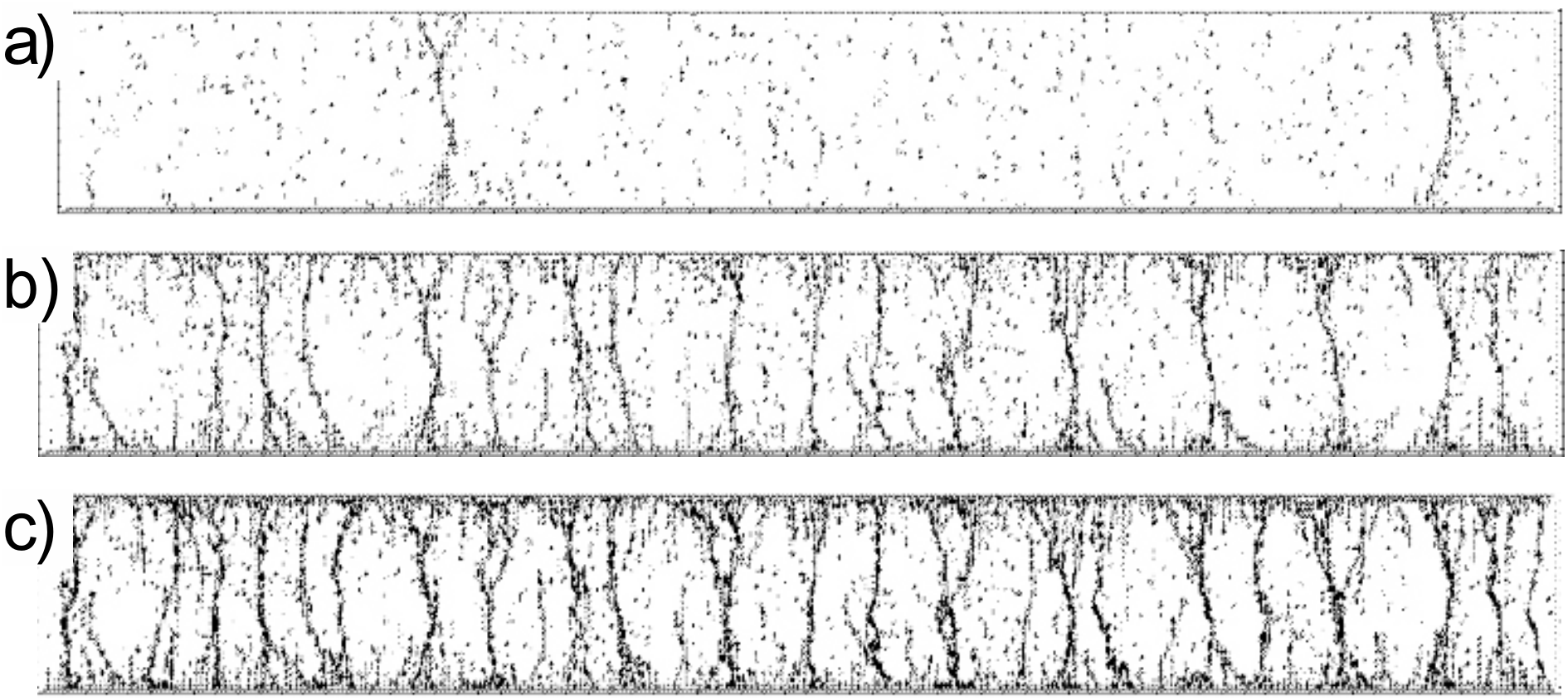}}
 \caption{ \label{fig:snapshots} Snapshots of the model system of the horizontal and vertical size $nx=800$ (cells per row) and $ny=30$ (number of rows). $(a)$ nucleated micro-cracks distributed over the  whole test cell, and the formation of the first segmentation crack. $(b)$ Quasi periodic segmentation pattern. $(c)$ Segmentation cracks and delaminations close to the crack density saturation state.}
\end{figure}

The discrete element method gives the possibility to monitor the microscopic damage development in the specimen. In accordance with the experiments simulations show, that the damage process of the $90^{\rm o}$ ply is composed of two distinct parts. At the beginning of loading the weakest  springs break in an uncorrelated fashion when they get over-stressed, generating micro-cracks in the specimen. This primary uncorrelated micro-crack nucleation in Fig.\ \ref{fig:snapshots}$a$ is dominated by the disorder of the system introduced for the damage thresholds of the springs. The micro-cracks substantially change the local stress distribution in the ply leading to stress concentrations around failed regions, which gives rise to correlated crack growths under further loading. Growing cracks or cracks formed by the coalescence of growing cracks can span the entire thickness of the $90^{\rm o}$ ply resulting in segmentation, {\it i.e.} break-up of the ply into pieces, as it is illustrated in Fig.\ \ref{fig:snapshots}$b$. Reaching the ply interface the crack stops without the possibility of penetration into the $0^{\rm o}$ ply in the model, as it was observed in the experiments \cite{schulte-fischedick-etall-99}, but with the possibility of deflection at the ply interface.
%%%%%%%%%%%%%%%%%%%%%%%%
%%%%%%%%%%%%%%% quer.eps 
Further segmentation cracks mainly form between existing cracks until the crack density is saturated due to occurring micro-delaminations. Micro-delaminations usually start to occur when the crack density has already reached high values, depending on the thickness of the transverse ply. Fig.\ \ref{fig:snapshots}$c$ shows that, as a result of the formation of spanning cracks, the ply gets segmented into pieces. The delamination zones along the interface can also be seen in Fig.\ \ref{fig:snapshots}$c$.%

\begin{figure}[htb]
 \centering{\includegraphics[scale=0.6]{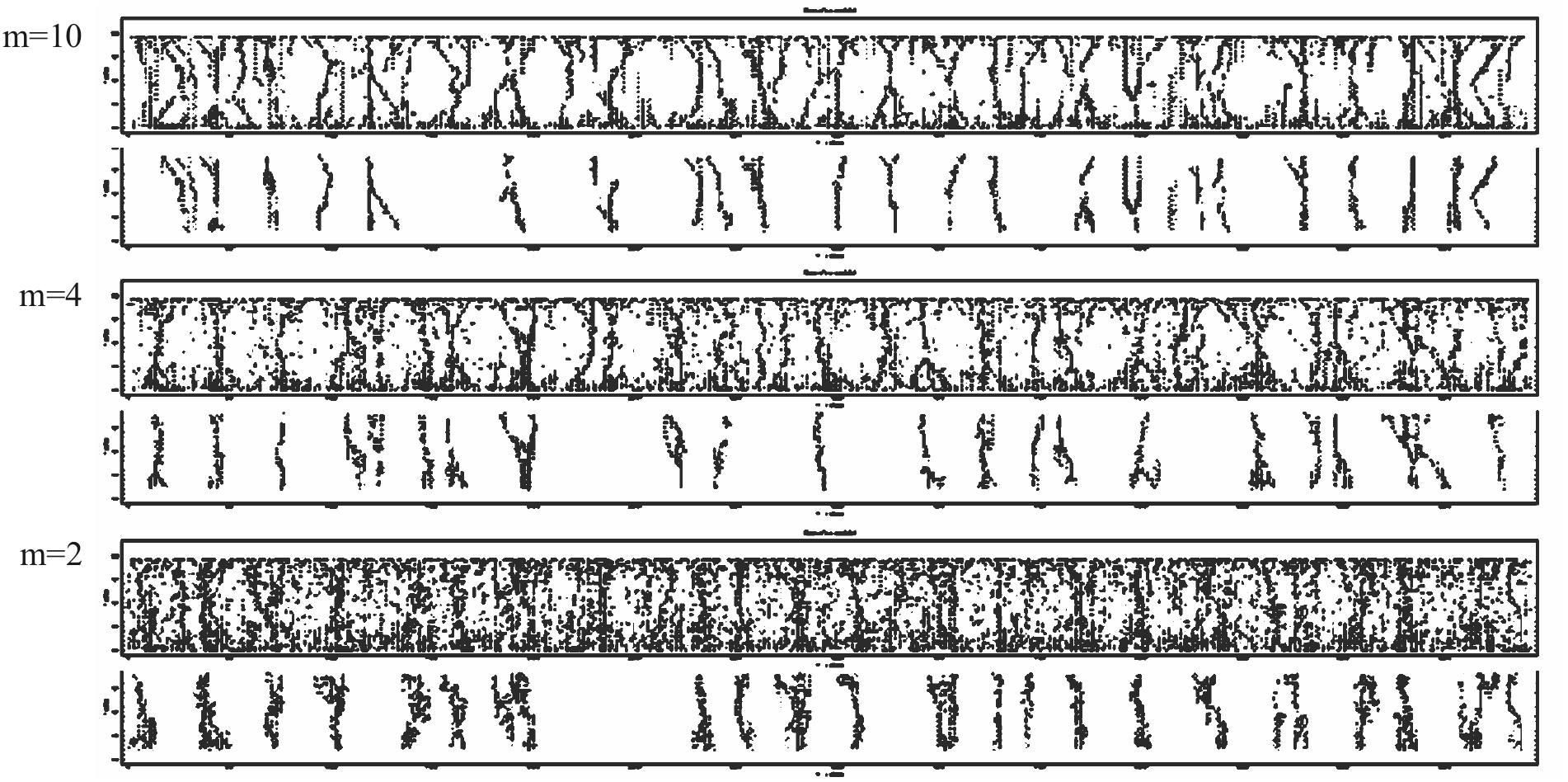}}
 \caption{ \label{fig:snapshots2}Snapshots of model systems of the size $nx=800$, $ny=20$ and identified clusters with disorder parameter $m$.}
\end{figure}

In order to obtain a quantitative characterization of the micro-structure of damage, we analyzed the spatial distribution of micro-cracks, {\it i.e.} the distribution of broken springs. A crack is identified as a connected set of broken springs in the triangular lattice taking into account solely nearest neighbor connections. A crack is considered to be a segmentation crack if it spans from one side of the $90^{\rm o}$ ply to the other. Since cracks forming along the interface of the plies complicate the identification of segmentation cracks in the framework of our algorithm, first micro-cracks along the interfaces are left out of the analysis. Identified clusters with this method are demonstrated in Fig.\ \ref{fig:snapshots2} for different values of disorder. The width of the clusters increases with the disorder, but still a good identification is possible even for large disorders. The method is illustrated in detail in \cite{wittel-kun-etal-2001}. It can be observed that the number of micro-cracks and also the number of segmentation cracks increases with the strain and the segmentation cracks have a more or less quasi-periodic spacing. The formation of the segmentation cracks is very fast and often with stable crack growth only during the crack initiation. This behavior is in good agreement with the experimental observations.
%%%%%%%%%%%%%%%%%%%%%%%%
%%%%%%%%%%%%%%% quer.eps 

\begin{figure}[htb]
 \centering{\includegraphics[scale=0.6]{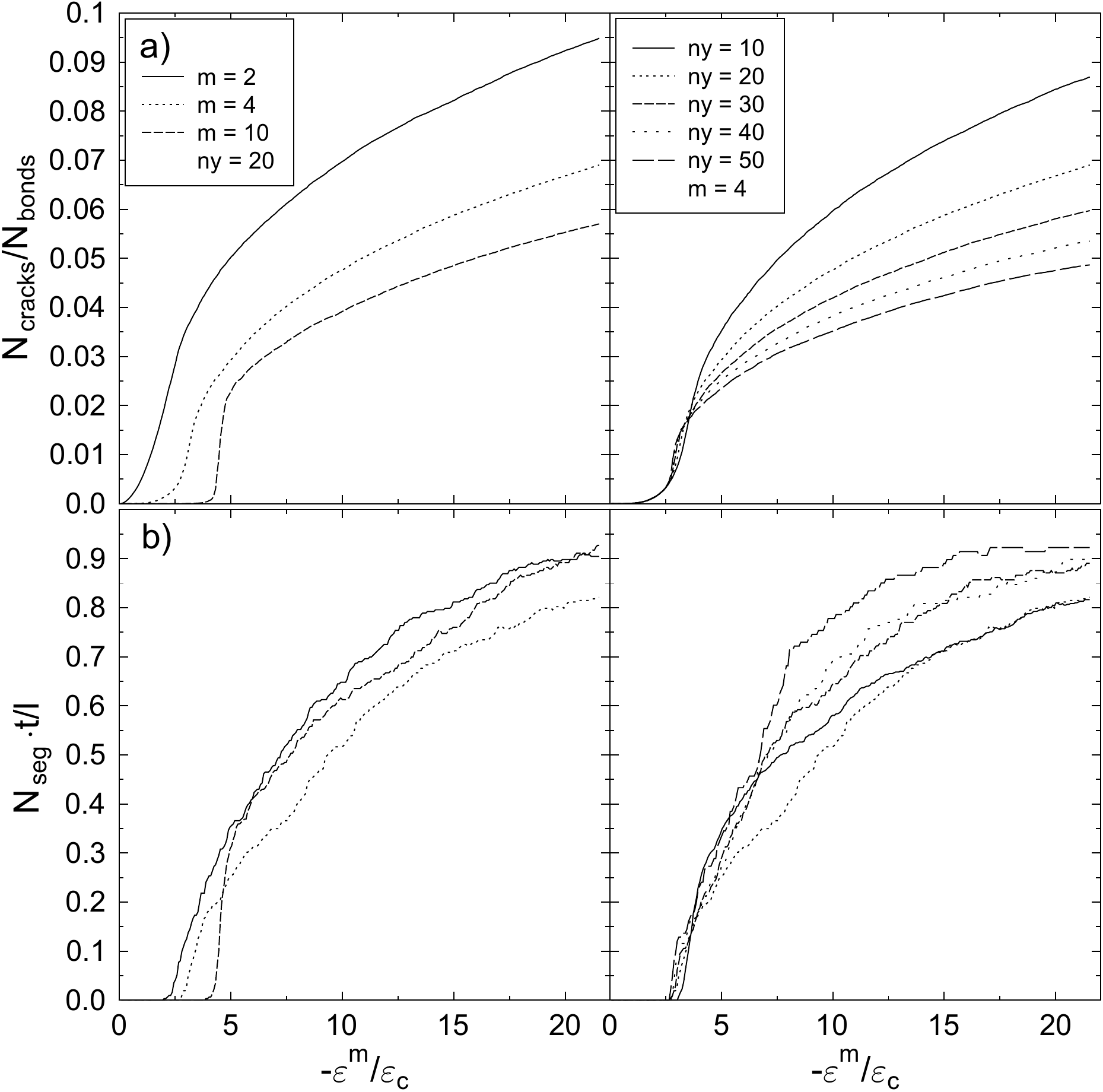}}
 \caption{\label{fig:crackdist} $(a)$ The total number of micro-cracks $N_{cracks}$ divided by the total number of bonds $N_{bonds}$, and $(b)$ the dimensionless crack density $N_{seg}\cdot t/l$ as a function of strain $\varepsilon^m/\varepsilon_c$. The left row shows results for constant ply thickness of 20 cells and varying disorder, while in the right row disorder is fixed to $m=4$ and ply thickness is varied. Smooth curves were obtained by averaging over nine samples.}
\end{figure}

The fraction of micro-cracks $N_{cracks}/N_{bonds}$ and the dimensionless segmentation crack density $N_{seg}\cdot t/l$ are presented in Fig.\ \ref{fig:crackdist} as a function of the strain $\varepsilon^{m}/\varepsilon_{c}$ for several different values of the Weibull-modulus $m$. $\varepsilon_c$ is the average critical strain for spring elements. Since $t$ is the ply thickness and $l$ the overall system length, $N_{seg}\cdot t/l$ can be understood as the thickness to width aspect ratio of segments. Varying $m$ can be interpreted as one major effect of the thermal fiber pretreatment: decreasing $m$ corresponds to a reduction of functional groups on the fiber surface, resulting in a clustering of functional groups along the fibers \cite{eckenhorst-mueller-etall-98} and consequently in larger disorder in the system. The second major effect is the reduction of the interface shear strength as the number of chemical bonds in the fiber-matrix interface is reduced.
%%%%%%%%%%%%%
%%%%%%%%%%%%  
It can be observed that cracking initiates at a finite strain value $\varepsilon_{in}$ called damage initiation strain, and starting from this point $N_{cracks}$ monotonically increases during the entire loading process. Segmentation first occurs at a strain larger than $\varepsilon_{in}$ after the number of micro-cracks reached a certain value. The number of segmentation cracks $N_{seg}\cdot t/l$ also increases, however, its value approaches a saturation value for larger strains in accordance with experimental observations.
 
Fig.\ \ref{fig:snapshots2} and \ref{fig:crackdist} provide an insight into the role of disorder in the damage process. Increasing the value of $m$, {\it i.e.} making the Weibull distribution narrower, the damage initiation strain $\varepsilon_{in}$ and the corresponding strain value for the start of segmentation increase, furthermore, the saturation value of the segmentation cracks increases. The slight decrease of $N_{seg}$ for wider breaking threshold distributions is an artifact of the cluster analysis. A clear identification of segments becomes more difficult as segmentation cracks, located close to each other, are connected via delaminations and cracks. For large disorder it is possible that they are identified as one cluster, and consequently as one segmentation.

\subsubsection{Micro-cracking in laminates with varying cross ply thickness}
In order to get an insight into the dependence of the micro-structure of damage on the model thickness, systems were simulated with thickness ranging from approximately $0.1$mm up to $0.5$mm. 
%%%%%%%%%%%%%%%%%%%%%%%%%
\begin{figure}[htb] 
 \centering{\includegraphics[scale=0.6]{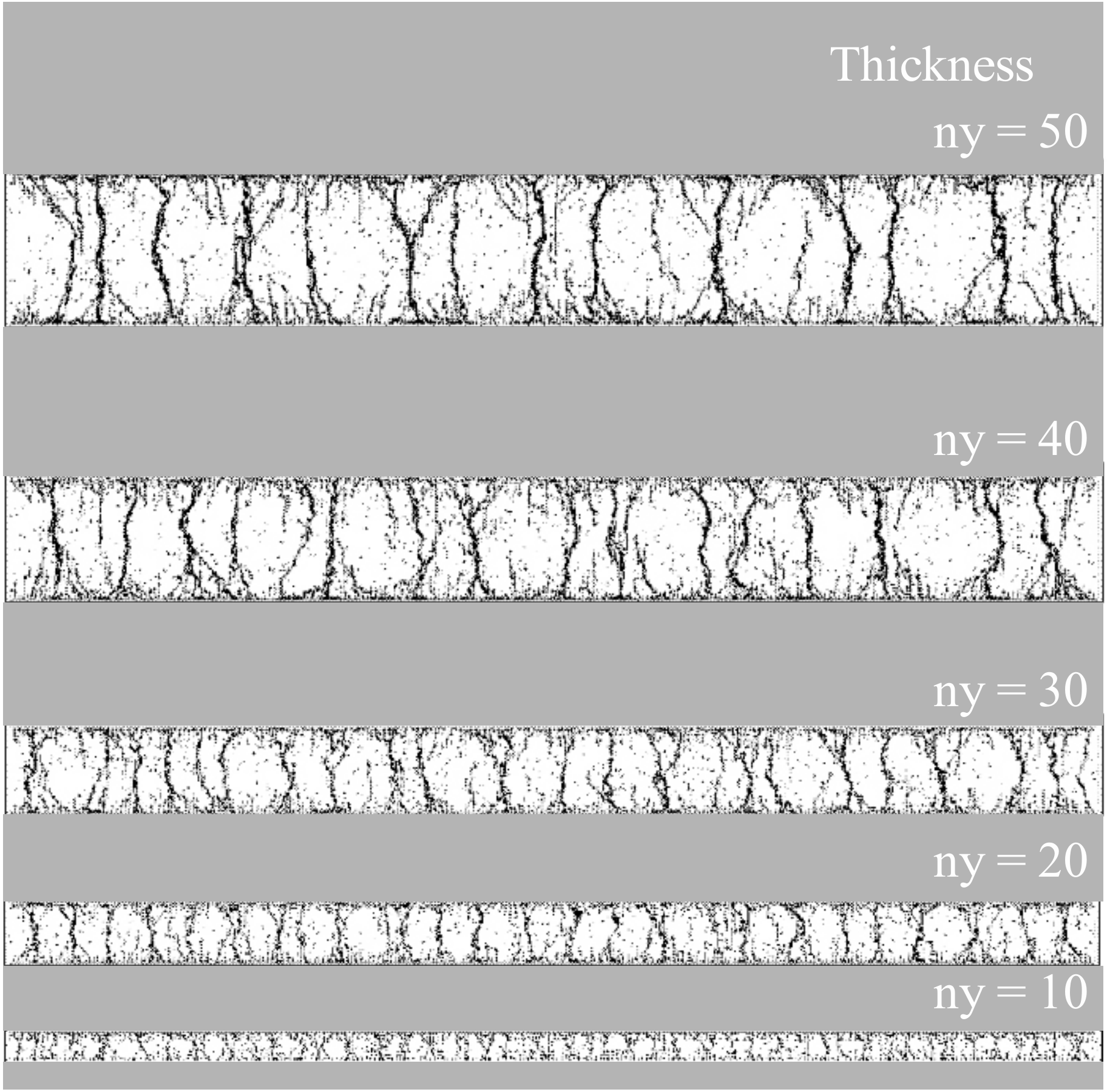}}
 \caption{\label{thickness} Micro-structure of damage for varying model thickness. The number of rows of cells $ny$ is given in the figure.}
\end{figure} 
%%%%%%%%%%%%%%%%%%%%%%%%%
The micro-structure of damage in Fig.\ \ref{thickness} shows a quasi-periodic spacing for all calculated thicknesses. For the thicker models branching of the segmentation cracks is observed close to the ply interface. This effect can complicate the cluster evaluation, since neighboring segmentation cracks can coalesce. Thinner systems reach their saturation crack densities at higher strains than the thick ones. These results are compared to an analytical approach and experiments in section \ref{analytical}.

\subsubsection{Development of the porosity}
The porosity of the pyrolysed material governs the constitutive behavior of the C/C-SiC material, since the infiltration with liquid silicon mainly depends on the accessibility of the crack systems from the outside. This part of the porosity is called open porosity, while the total porosity is defined as the total volume of cracks, including those located deep inside the material without any connection to the outside. Therefore, all elementary cracks that are not connected to the outside have to be identified in a cluster analysis. These isolated cracks are used to calculate the area of the closed porosity $A^{cc}$, by adding the crack area of all broken elements forming theses cracks. The crack area between element $i$ and $j$ is approximated with equilateral triangles, computing the difference between the actual area and the uncracked, stress free area at this temperature,
\begin{equation}
       	A^{cc}_{ij}(T)=\frac{1}{2\sqrt{3}}\cdot\left({r_{ij}^2(T)-l^{0^2}(T)}\right).
\end{equation}
Since segments delaminate from the ply interface in the course of the simulation, the open porosity can be evaluated by using the difference of the areas 
\begin{equation}
	A_{oc}(T)=A(T) - A^0(T) - A^{cc}(T),
\end{equation}
with $A^0(T)$ being the reference value of the stress free system without any damage and consequently with a porosity of 0\%. $A^{cc}$ is the sum of the area of all broken elements $ij$ of all identified isolated cracks.
%%%%%%%%%%%%%%%%%%%%%%%%%
\begin{figure}[htb] 
  \centering{\includegraphics[scale=0.95]{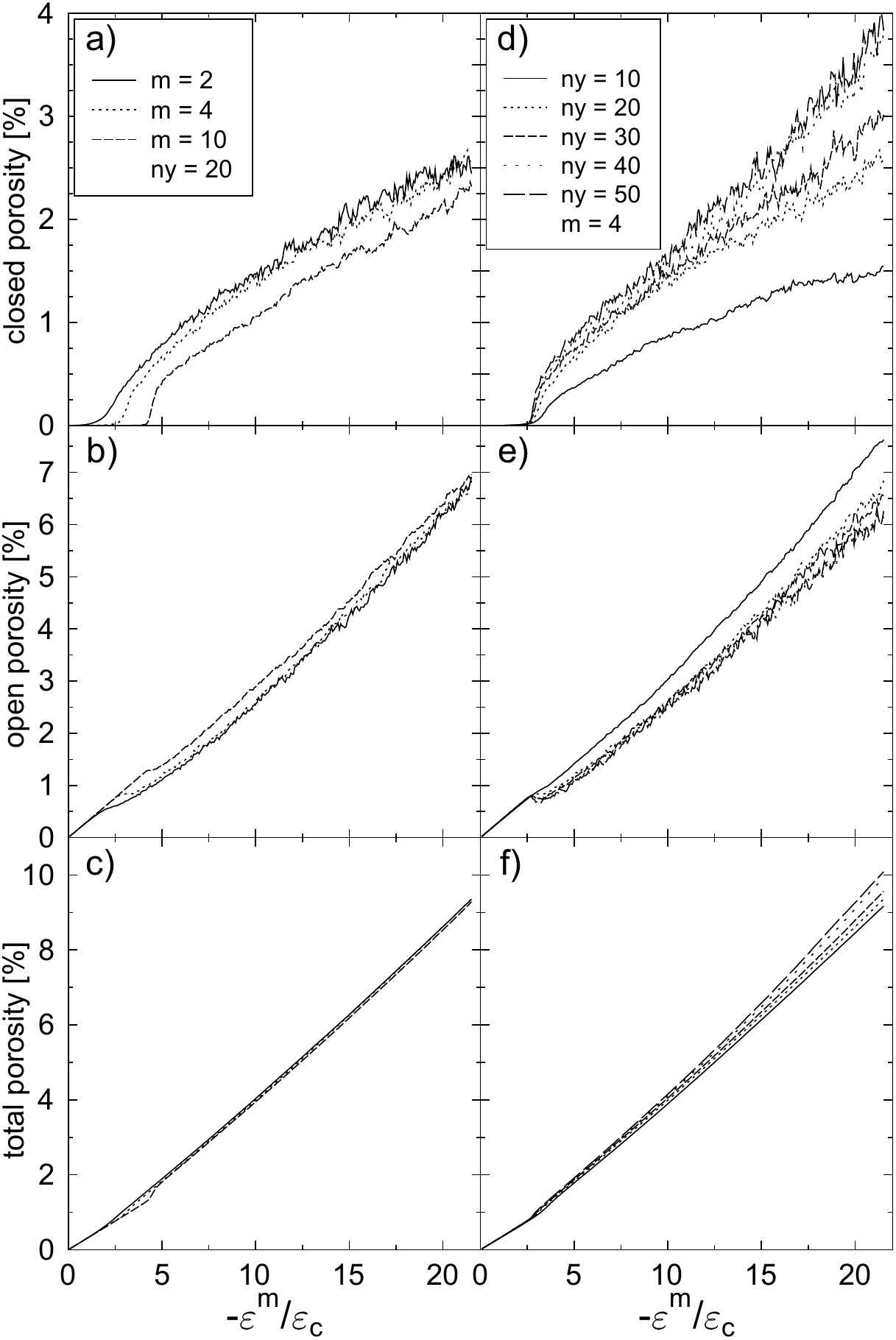}}
  \caption{\label{porosity} Development of the porosity for varying disorder and constant model thickness $ny=20$ ($a-c$) and varying model thickness but constant disorder $m=4$ ($d-$) respectively.}
\end{figure} 
%%%%%%%%%%%%%%%%%%%%%%%%%
The total area of the model $A(T)$ is calculated from the positions of all edge elements. It is obvious that the open porosity calculated before stress can be released by forming cracks is artificial, but as soon as crack patterns form, the assumptions for the calculations become realistic. Nevertheless, the open porosity from measurements and from optical area analysis (see Fig.\ \ref{porosity_crackdensity}) is approximately 13 \% and therefore larger than the calculated ones that are 6-7\% for this case. Even though quantitative agreement is beyond the scope of the model, the effects of disorder and ply thickness on the porosity are obtained correctly. Note that the two-dimensional model, which represents a small zone inside the material, is limited to its characteristic scale and can not take effects into account, that are caused by the third dimension. A pyrolysed resin sample as well as a fiber bundle, therefore, are calculated with the porosity equal to $0\%$. On the other hand, the real material is irregular with resin rich zones, small delamination zones and a large specimen surface. It is also possible, that water intrudes the small cracks and gas channels along the fibers below the model scale. The dependence of the total porosity on the disorder of the system in Fig.\ \ref{porosity}$c$ is very small, but for the system with low disorder, the closed porosity (Fig.\ \ref{porosity}$a$) is smaller than for the highly disordered system. Even though thick systems have a much smaller number of segmentations cracks than thinner systems, the open porosity in Fig.\ \ref{porosity}$e$ does not show a strong dependence on the system thickness. This is mainly due to increasing crack openings in thicker systems. The closed porosity shows a stronger dependence on the thickness, with increasing values for increasing system sizes. It is therefore responsible for the deviation of the total porosity in Fig.\ \ref{porosity}$f$ from a master curve.

\section{Analytical model} \label{analytical}
The micro-mechanics of damage (MMD) analysis, used here for comparative reasons, is described in detail in \cite{nairn-2000} and was employed in \cite{wittel-kun-etal-2001}. A MMD analysis is based on the individual prediction of the initiation and propagation of the various types of damage, including possible interactions between all identified relevant damage mechanisms. The analysis consists of two steps, {\em (i)}~a stress analysis, and {\em (ii)}~a damage analysis for the expected damage mechanisms. The stress analysis has to be undertaken in the presence of observed damage and is completely independent from the postulate of a failure criterion for the initiation and evolution of damage, but of course not from the damage itself.   

\begin{figure}[htb] 
 \centering{\includegraphics[scale=0.95]{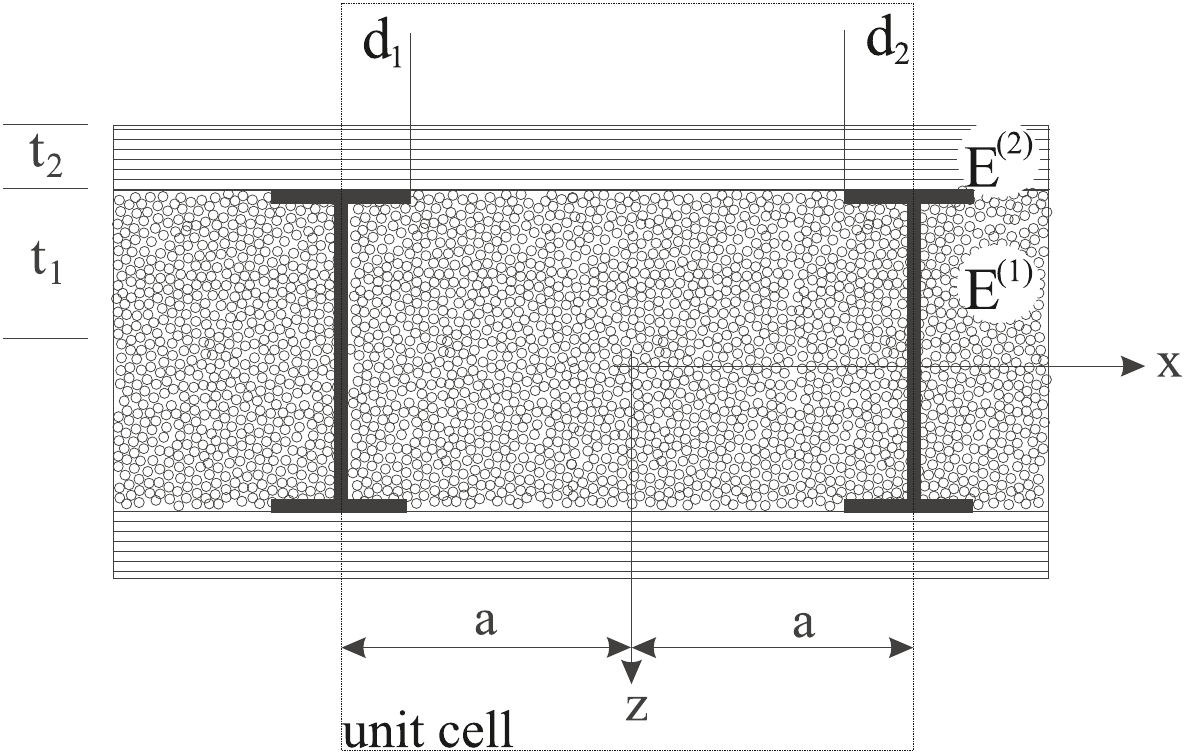}}
 \caption{\label{cdtransdef} Side view on a unit cell representing a cross ply laminate with segmentation cracks ($a$=half crack spacing) and delaminations $d_1,d_2$ in the transverse ly with half thickness $t_1$.} 
\end{figure}

{\em (i) Stress analysis:}
The stress field in the presence of micro-cracking and delamination damage is calculated for a unit cell bounded by cracks located at each end (Fig.\ \ref{cdtransdef}). The entire specimen is built by a sequential formation of unit cells. A two-dimensional analysis for the $x-z-$plane derived in Ref.\ \cite{hashin-85} using a variational mechanics analysis is employed. Following  \cite{hashin-85} we assume that the $x$-axis tensile stress in each ply group is independent of $z$, that cracks span the entire ply thickness and that delaminations are always symmetric to the crack plane\cite{nairn-2000}.
%%%RVE-Bild%%%%%%%%%% 
%%%%%%%%%%%%%%%%%%%%%%%%% 
The x-axis tensile stress of a micro-cracked laminate can be derived with force balance 
\begin{equation}\label{sigmaxx} 
 \sigma^{(1)}_{xx}=k^{(1)}_m\sigma_0-\psi(x) \qquad 
 \sigma^{(2)}_{xx}=k_m^{(2)}\sigma_0+\frac{\psi(x)}{\lambda} 
\end{equation} 
with the undetermined function $\psi(x)$, the applied global stress $\sigma_0$ and $\lambda=t_2/t_1$, expressing stress perturbations caused by the segmentation cracks. The integration of the stress equilibrium equations with the unit cell boundary conditions gives the shear and transverse stresses in the unit cell in terms of $\psi(x)$. The function $\psi(x)$ that minimizes the complementary energy gives the best approximation to the micro-cracked cross-ply laminate stress state. The solution of the Euler equation for finding $\psi(x)$ is dependent on the temperature and can be found in detail in \cite{hashin-85}.

{\em (ii) Damage analysis:} The failure criterion used, is an energy criterion, allowing a formation of a new segmentation crack, whenever the strain energy release rate (ERR) $G_m$ associated to the formation of a new micro-crack exceeds some critical value $G_{mc}$, also called intra-laminar fracture toughness or micro-cracking fracture toughness. The ERR is also calculated with the two-dimensional, variational mechanics analysis \cite{nairn-2000}.
 Micro-crack induced delamination competes with the formation of new micro-cracks. Experimental and numerical studies show, that micro-crack induced delamination does not start before a critical micro-crack density is reached.

\begin{figure}[htb] 
 \centering{\includegraphics[scale=0.5]{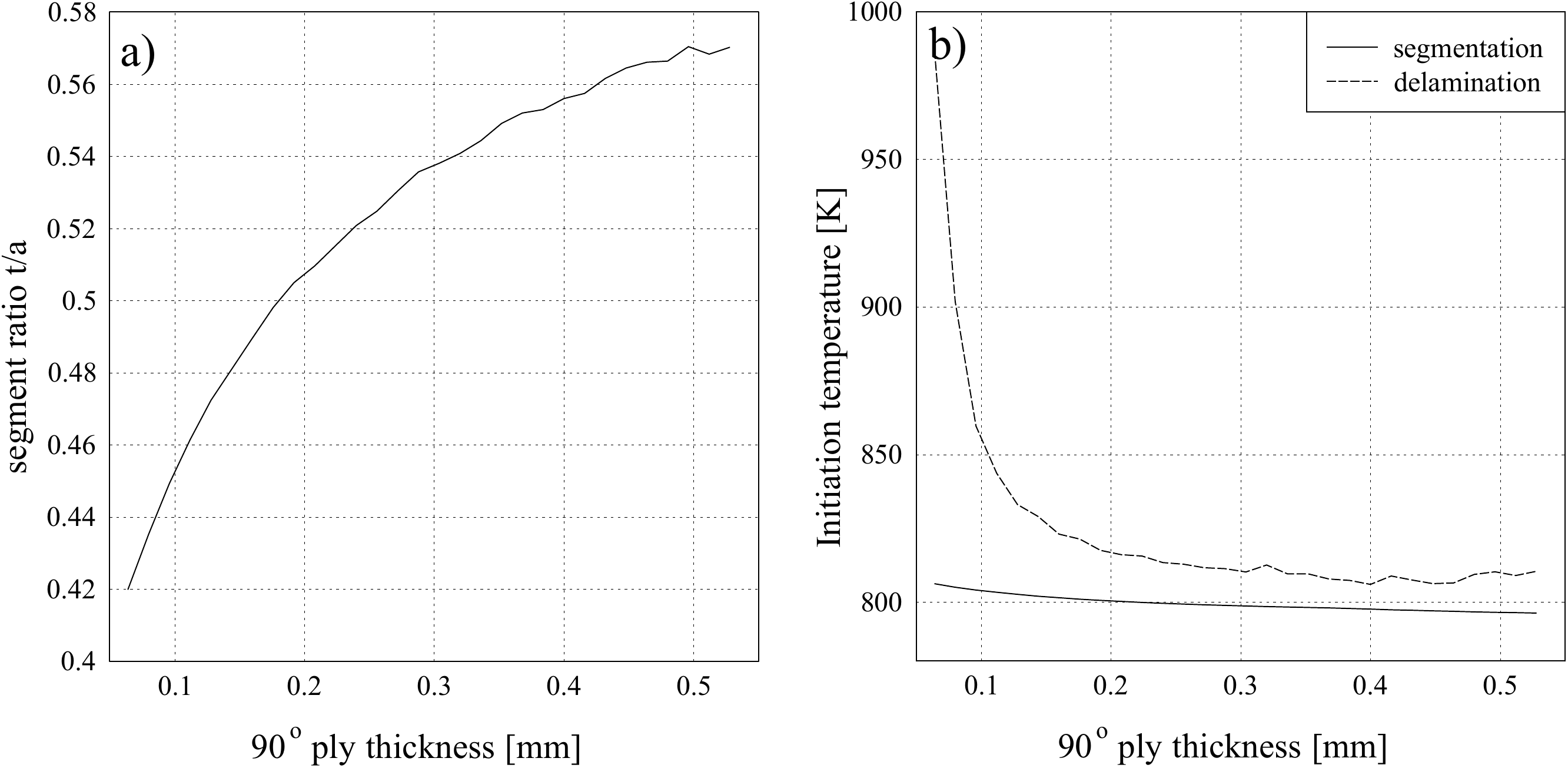}}
 \caption{\label{3dplot} $a)$ Segment thickness/width ratio for saturation and $b)$~initiation temperatures for segmentation and delamination as a function of the ply thickness} 
\end{figure}

The two-dimensional, variational mechanics analysis described above can be extended to account for delaminations emanating from micro-crack tips. The assumption that $\sigma_{xx}^{(1)}$ is independent of $z$, indicates identical delamination lengths emanating from the top and bottom micro-crack tips. Following \cite{nairn-2000} we furthermore assume that $G_{mc}$ and the critical ERR associated with the growth of delamination $G_{dc}$ are equal. This way, the dimensionless crack saturation density in Fig.\ \ref{3dplot} was calculated. In principal, it is possible to get better agreement between MMD analysis and the experimental data by adapting $G_{mc}$ and $G_{dc}$ until a satisfying fit is achieved, since these values have an ample scope for physical interpretation. Nevertheless, we find qualitative agreement in the increase of $t/a$ as a function to the $90^\circ$ ply thickness for numerical and analytical results (compare Figs.\ \ref{3dplot} and \ \ref{fig:crackdist}).

{\em Thermal ply degradation with MMD:} During the carbonization process, the shrinkage is opposed to the thermal expansion of the material. Therefore, experimental data on the strain-temperature behavior of the material from Fig.\ \ref{epsilon_ud} is required.
In-situ experiments on the development of material properties like Young's modulus $E_x,E_z$ or material strength during the carbonization process without influencing the process of crack formation is believed to be beyond todays possibilities. We use the experimentally evaluated laminate properties $E_{xUD}=0.0235\cdot T+138.9$GPa, $E_{zUD}=0.00917\cdot T+3.93$GPa and $\nu_{xzUD}=166.17\cdot T^{-1}+0.206$ for  $T>400$K, measured at room temperature after incremental pyrolysis as well as the values for $G_{yzUD}=3.36$GPa, $G_{xzUD}=6.25$GPa, $\nu_{yzUD}=0.38$, $G_{mc}=55.7$J/m$^2$, evaluated at room temperature.
%%%%%%%%%%%%%%%%%%%%%%%%% 
%%%%%%%%%%%%%%%%%%%%%%%%% 

The behavior for segmentation initiation and crack saturation shown in Fig.\ \ref{3dplot} was observed in the numeric calculations. For very thin plies, delamination does not occur within the temperature range. Transferred to laminate of fabric this means, that the onset of micro delaminations preferentially occur at the thicker regions and are stopped at the tip of the cross-sectioned fiber bundle. This effect explains why micro delaminations normally do not extend to macro delaminations.

For a comparison of experimental with numeric and analytic results, we use a dimensionless strain $\varepsilon=\varepsilon_0 / \varepsilon_c \quad \varepsilon_c=\varepsilon_{B,resin}=1.0176$ in order to fit the data to the temperature curve.
%%%%%%%%%%%%%%%%%%%%%%%
\begin{figure}[htb] 
 \centering{\includegraphics[scale=0.7]{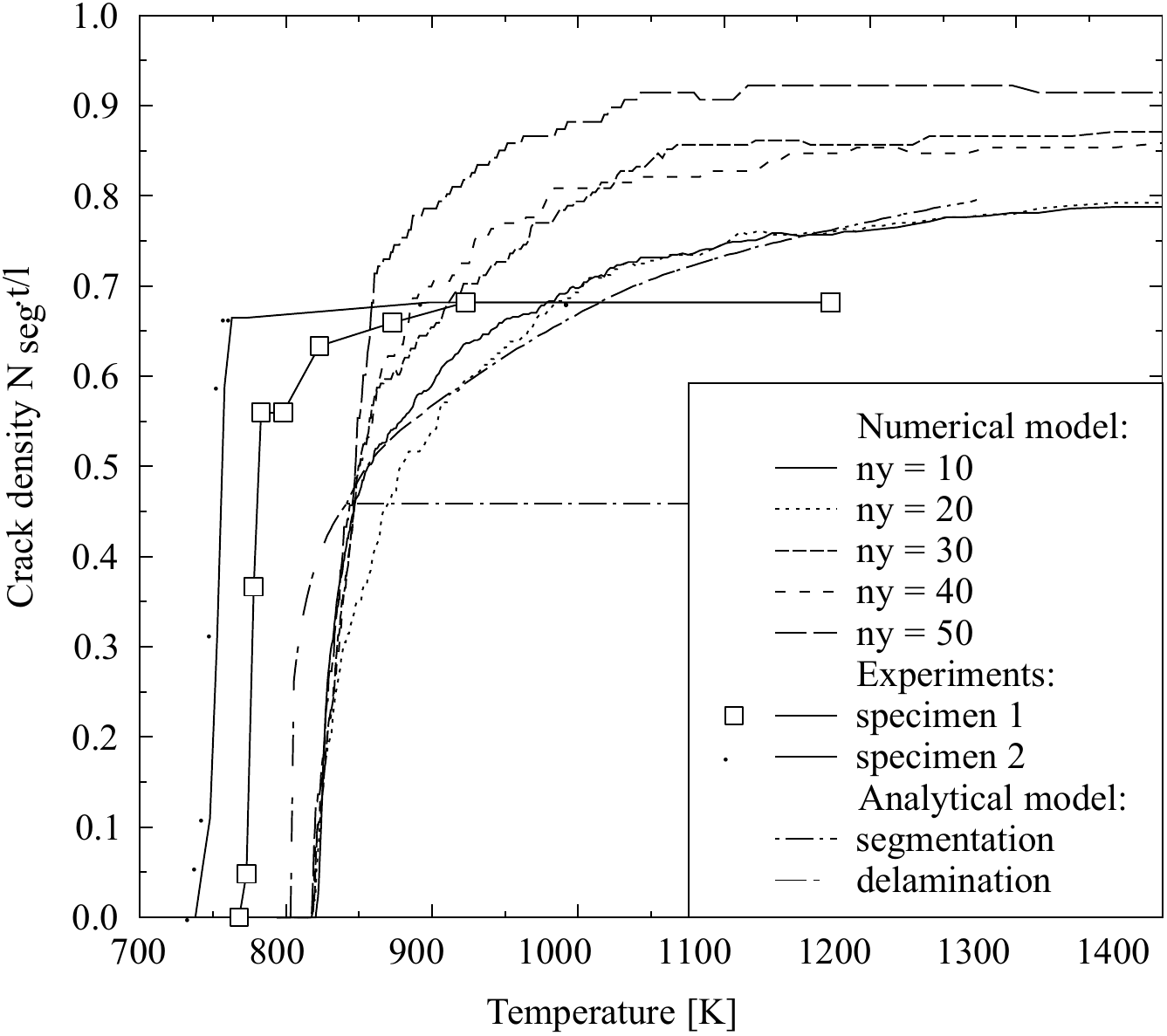}}
 \caption{\label{comparison} Comparison of numerical and analytical approaches with experimental data.}
\end{figure}
%%%%%%%%%%%%%%%%%%%%%%%
Cracks were counted only if they span the entire ply and if they are clearly visible, using a microscope. Therefore, the measured crack density of 5.98 1/mm (equivalent to $N_{seg}\cdot t/l=0.68$) has to be considered as a lower bound value. Note that the initiation temperatures are only from two experiments and show a large scatter. The analytic as well as the numeric approach qualitatively capture the fast increase of the dimensionless segmentation crack density or segment ratio and the saturation in Fig.\ \ref{comparison}. Reasonable agreement is achieved between the numeric approach and the crack saturation value of the experiments (ny=10-20). The analytic solution for the saturation density is around 0.45 and far from the experimental value of 0.68. This is mainly due to the model assumption $G_{mc}=G_{dc}$ and experimental problems of evaluating these values. Nevertheless, the analytic solution provides a qualitative insight on the role of ply thickness on initiation values (comp. Fig.\ \ref{3dplot}). Note that the segment width in the analytical model is defined as $2a-d_1-d_2$. The assumption of nearly constant material properties and no creep is believed to be responsible for the difference in the initiation temperature from the experiments. 

\section{Conclusions}
We introduced a disordered spring network model to study the transverse cracking of the $90^\circ$ ply in $[0/90]_S$ cross-ply laminates, applied for the case of thermal degradation of the matrix material during the pyrolysis process. The main advantage of our modeling is that it naturally accounts for the complicated local stress fields formed around failed regions in the material, furthermore, it captures the gradual activation of the relevant failure mechanisms and their interactions during the fracture process. We have demonstrated that our discrete element model provides an insight into the damage process occurring under gradual matrix degradation of cross ply laminates. Quantitative results have been obtained on the micro-structure of damage and on the development of the porosity for varying ply thickness and disorder.

The results obtained by numerical simulations have been confronted with the experimental findings and also with an analytic approach. Reasonable agreement was found between the numerical, analytical and experimental results. However, the numerical simulations proved to be more realistic than the simple analytic approaches, due to the more detailed description of micro-structure in the discrete element model.

The results of the simulation support material optimization efforts. It was found that with decreasing ply thickness, the tendency for the development of delaminations is clearly reduced (see Fig.\ \ref{3dplot}$b$). Applied to CMC derived from woven fabrics, this could be achieved by reducing the number of filaments per roving, the so called K-number. This result gains importance as the development of delaminations is the size limiting factor in pyrolytically produced CMC. Thus, larger components may be produced by using finer woven laminates. On the other hand, this leads to a higher open porosity (see Fig.\ \ref{porosity}) and in the case of the LSI-process to a higher amount of fiber degradation during siliconization. 
The increase in open porosity with decreasing ply thickness can be utilized in producing functionally graded CMCs by the LSI-process. A single 1K-laminate on top of the standard C/C-SiC would give a gradual transition for the SiC-content from the inner parts to the totally converted surface of the material. This could give a better adherence of coatings like environmental barrier coatings to the substrate. Based on predictions for the development of crack morphology, porosity and delamination behavior for different system sizes and disorder, promising approaches for new materials have been proposed.

\section{Acknowledgment} 
The presented work is partly funded by the German Science Foundation (DFG) within the Collaborative Research Center SFB 381 'Characterization of Damage Development in Composite Materials using Non-Destructive Test Methods' which is gratefully acknowledged, as well as the support under NATO grant PST.CLG.977311. F.\ Kun is grateful for financial support of the Alexander von Humboldt Stiftung (Roman Herzog Fellowship), and also for the B\'olyai J\'anos fellowship of the Hungarian Academy of Sciences and for support of the research contract FKFP0118/2001. J. Schulte-Fischedick is grateful for the support within the Graduate Program (GRK 285) 'Interfaces in Crystalline Materials'.

\end{document}